\documentclass[aps,prd,preprint,showpacs]{revtex4}

\usepackage{longtable}
\usepackage{textcomp}

\begin{document}
\title{Intrinsic light and strange quark--antiquark pairs in the proton and nonperturbative
strangeness suppression}
\author{C. S. An$^{1}$}\email{ancs@swu.edu.cn}

\author{B. Saghai$^{2}$}\email{bijan.saghai@cea.fr}

\affiliation{1. School of Physical Science and Technology, Southwest University,
              Chongqing 400715, People's Republic of China\\
2. IRFU, CEA, Universit\'e Paris--Saclay, 91191 Gif--sur--Yvette, France}

\thispagestyle{empty}

\date{\today}

\begin{abstract}
%
%
The CLAS Collaboration recently reported measured ratios of pion and kaon electroproduction cross sections
from a proton target and extracted the ratios for light and strange quark--antiquark pairs,
${u\bar{u}}/ {d\bar{d}}$ and  ${s\bar{s}}/ {d\bar{d}}$.
Within an extended chiral constituent quark formalism, we investigate contributions to those ratios  from the
nonperturbative mechanism due to all possible intrinsic $|uudQ \bar{Q}\rangle$ Fock states in the proton;
with $Q \bar{Q} \equiv u\bar{u},~ d\bar{d},~s\bar{s},~c\bar{c}$.
Our results are compared with the CLAS data and findings from other phenomenological approaches, offering insights
into the manifestations of the genuine five--quark Fock states in the proton and its relevance to interpreting the
experimental  data.

\end{abstract}


\maketitle

%
\section {Introduction}
Flavor content of the nucleon is known to be an important issue in understanding the hadronization process and the
interactions of quarks in quantum chromodynamics (QCD).
Moreover, the intrinsic quark--antiquark components $|qqqQ \bar{Q}\rangle$ in the baryon wave functions are a prediction
of QCD and has been  under study since four decades (see; e.g., review papers
~\cite{Vogt:2000sk,Garvey:2001yq,Chang:2014jba,Brodsky:2015fna} and references therein),

With respect to the nonperturbative mechanisms, the proton light flavor asymmetry
$\mathcal{A}_p=\bar{d} - \bar{u}$ and the entity $ \bar{u}  + \bar{d} - s - \bar{s}$,
are of paramount interest, given that they are 
free from the contributions of the extrinsic sea quarks (e.g. gluon splitting $g \to{Q} \bar{Q}$) ~\cite{Chang:2014jba}.
However, as discussed in Sec.~\ref{sec:Rratio}, the present experimental knowledge does not allow us putting sharp
enough constraints on the phenomenological models.

A new piece of information on the quark--antiquark pairs was released by the CLAS Collaboration ~\cite{Park:2014zra}
on the ratios  ${u\bar{u}}/ {d\bar{d}}$ and ${s\bar{s}}/ {d\bar{d}}$, which were recently interpreted by
Santopinto~{\it et al.} \cite{Santopinto:2016fgs} within the unquenched quark model (UQM).
Those ratios can also be extracted from other approaches.
Actually, Chang and Peng~\cite{Chang:2011a,Chang:2011b,Chang:2014lea} investigated the  intrinsic ${Q\bar{Q}} $ states in the
proton by generalizing to the light and strange quark--antiquark pair components the pioneering work on the intrinsic sea  $|uudc \bar{c}\rangle$
by Brodsky, Hoyer, Peterson, and Sakai~\cite{Brodsky:1980pb}, the BHPS model.
Moreover, the  Neural Networks for Parton Distribution Functions (NNPDF) Collaboration~\cite{Ball:2012cx} determined the ratios of
the strange quark--antiquark pairs  to those of light ones, coming from both intrinsic and extrinsic contributions.

The goal of the present study is to predict the contributions to those ratios arising from the intrinsic Fock states
in the proton's wave function.
Our formalism~\cite{An:2012kj} is based on the extended chiral constituent quark approach and embodies all possible five--quark
mixtures in the proton's wave function;  with the mechanism of transition between three- and five--quark components in the proton
treated within the $^{3}P_{0}$ quark--antiquark creation frame~\cite{Le Yaouanc:1972ae,Le Yaouanc:1973xz,Kokoski:1985is}.

The present manuscript is organized in the following way:
In Sec.~\ref{sec:recall} we introduce the  theoretical frame and give explicit expressions for the probabilities of the intrinsic
quark--antiquark pairs ($P_{p}^{Q\bar{Q}}$) in the proton (Sec.~\ref{sec:proba});  relating them to the studied
ratios, namely, ${u\bar{u}} / {d\bar{d}}$ , ${s\bar{s}} / {d\bar{d}}$, and
$2{s \bar{s}} / ({u\bar{u}} + {d \bar{d}})$.
In Sec.~\ref{sec:Res} we present our  numerical results for $P_{p}^{Q\bar{Q}}$~(Sec. \ref{sec:Rproba}) and
for the quark-antiquark ratios.
Comparisons with the data and the outcomes from other approaches
~\cite{Santopinto:2016fgs,Chang:2014lea,Ball:2012cx} are reported in (Sec.~\ref{sec:Rratio}).
Finally, Sec.~\ref{sec:conclu} is devoted to a summary and conclusions.
%
%
\section{Theoretical frame}
\label{sec:Theo}
The content of our extended chiral constituent quark model (E$\chi$CQM) was developed
in~\cite{An:2012kj,An:2013daa,An:2014aea,Duan:2016rkr}. Hence, in Sec.~\ref{sec:recall}, we briefly present the
main features of the formalism. In Sec.~\ref{sec:proba} we give explicit expressions for the light
and strange quark--antiquark pair probabilities in the proton and relate them to the studied ratios.
\subsection{Extended chiral constituent quark approach}
\label{sec:recall}
The wave function for the baryon $B$ can be written in the following form:
\begin{equation}
 |\psi\rangle_{B}=\frac{1}{\mathcal{\sqrt{N}}}{\Big[}|qqq\rangle+
 \sum_{i,n_{r},l}C_{in_{r}l}|qqq(Q \bar{Q}),i,n_{r},l\rangle {\Big ]}\,,
\label{wfn}
\end{equation}
where the first term is the conventional wave function for the baryon with three constituent quarks  and the
second term is a sum over all possible higher Fock components with a $Q \bar{Q}$ pair;
$Q \bar{Q} \equiv u \bar{u},~d \bar{d},~s \bar{s},~c \bar{c}$~\cite{An:2012kj,Duan:2016rkr}.
Different possible orbital-flavor-spin-color configurations of the four--quark subsystems in the five--quark system are numbered by
$i$, $n_{r}$, and $l$, denoting the inner radial and orbital quantum numbers, respectively. $C_{in_{r}l}/\sqrt{\mathcal{N}}$
represents the probability amplitude for the corresponding five--quark component.

The coefficient $C_{in_{r}l}$ for a given five--quark component can be related to the transition matrix element  between the  three-
and five--quark configurations of the studied baryon
\begin{equation}
C_{in_{r}l}=\frac{\langle qqq|  \hat{T}  |qqq(Q \bar{Q}),i,n_{r},l\rangle}{M_B-E_{in_{r}l}}\,,
\label{coef}
\end{equation}
where $M_B$ is the physical mass of baryon $B$ and $E_{in_{r}l}$ the energy of the five--quark component.

To calculate the corresponding transition matrix element,
we use a $^{3}P_{0}$ version for the transition coupling operator $\hat{T}$~\cite{RSB,Santopinto:2010zza}
\begin{eqnarray}
 \hat{T}&=&-\gamma\sum_{j}\mathcal{F}_{j,5}^{00}\mathcal{C}_{j,5}^{00}C_{OFSC}
 \sum_{m} \langle1,m;1,-m|00\rangle\chi^{1,m}_{j,5}\nonumber \\
&&\mathcal{Y}^{1,-m}_{j,5} (\vec{p}_{j}-\vec{p}_{5})b^{\dag}(\vec{p}_{j})d^{\dag}(\vec{p}_{5})\,.
\label{op}
\end{eqnarray}
In the above equation, $ \hat{T}$ has units of energy, so that  $\gamma$ is (in natural units)  a dimensionless constant of the model.
$\mathcal{F}_{i,5}^{00}$ and $\mathcal{C}_{i,5}^{00}$ are the flavor and color singlet of the quark--antiquark pair $Q_{i} \bar{Q}$
in the five--quark system, and $C_{OFSC}$ is an operator to calculate the orbital-flavor-spin-color overlap between the residual
three-quark configuration in the five--quark system and the valence three--quark system.
$\chi^{1,m}_{j,5}$  is a spin triplet wave function with spin $S$=1 and  $ \mathcal{Y}^{1,-m}_{j,5}$ is a solid spherical harmonics
referring to the quark and antiquark  in a relative $P-$wave.
$b^{\dag}(\vec{p}_{j})$ and $d^{\dag}(\vec{p}_{5})$ are the creation operators for a quark and antiquark with momenta
$\vec{p}_{j}$ and $\vec{p}_{5}$, respectively.
The operator $ \hat{T}$,  expressed in second-quantization form, can then be applied in the Fock space.

As reported in~\cite{An:2012kj}, out of 34 possible five--quark configurations, only 17 of them survive with nonvanishing
transition matrix elements; with orbital and radial quantum numbers $l$~=~1 and $n_r$~=~0, respectively.

The probability of the sea quark--antiquark pairs in the baryon $B$ and the normalization
factor read, respectively,
\begin{eqnarray}
\mathcal{P}_B^{Q \bar{Q}}&~=~& \frac{1}{\mathcal{N}}
\sum_{i=1}^{17}\Bigg[ \Big ( \frac{T_i^{Q \bar{Q}} }{M_B - E_i^{Q \bar{Q}}} \Big )^2 \Bigg ],
\label{proba}  \\ [10pt]
\mathcal{N}  &\equiv&  1+ \sum_{i=1}^{17} \mathcal{N}_i
=1+\sum_{i=1}^{17} \sum_{Q \bar{Q}} \Bigg[ \Big ( \frac{T_i^{Q \bar{Q}}}{M_B-E_i^{Q \bar{Q}}}\Big )^2  \Bigg ].
\label{norm}
\end{eqnarray}
Here $T_i^{Q \bar{Q}}$ denotes the transition matrix element of the $^{3}P_{0}$ operator in Eq.~(\ref{op}) between the 
$i$th five-quark component and the valence three--quark nucleon state, and $E_i^{Q \bar{Q}}$ the energy of the $i$th five-quark component.
The first term in Eq.~(\ref{norm})~is due to the valence three--quark state, while the second term comes from the five--quark
mixtures,  with $Q \bar{Q} \equiv u \bar{u},~d \bar{d},~s \bar{s},~c \bar{c}$.

%
%
\subsection{Quark-antiquark pair probabilities and ratios}
\label{sec:proba}
In this section, starting from Eq.~(\ref{proba}), we give explicit expressions for the light and strange quark--antiquark
pair ($u \bar{u}$, $d \bar{d}$, $s \bar{s}$) probabilities for the proton ($\mathcal{P}_p^{Q \bar{Q}})$
in terms of the five--quark probabilities per configuration [$P_p(i)$, $i$~=~1--17].

The probability amplitudes are calculated within the most commonly
accepted $Q\bar{Q}$ pair creation mechanism, namely, the $^{3}P_{0}$ model.
Then, the $Q\bar{Q}$ pair is created anywhere in space with the quantum numbers of the
QCD vacuum $J^{PC} = 0^{++}$, corresponding to $^{3}P_{0}$~\cite{Le Yaouanc:1972ae}.
This model has been successfully applied to the decay of mesons and baryons
~\cite{Le Yaouanc:1973xz,Kokoski:1985is}, and has recently been employed to analyze the
sea flavor content of the ground states of the $SU(3)$ octet baryons~\cite{An:2012kj} by
taking into account the $SU(3)$ symmetry breaking effects.

The probabilities of light quark--antiquark pairs for the proton in terms of the
relevant configurations [$P_p (i)$] and the  associated squared Clebsch-Gordan coefficients read
\begin{eqnarray}
\mathcal{P}_p^{u \bar u} &=& \frac{2}{3} \Big{[}P_p(3)+P_p(5)+P_p(9)+P_p(12)+P_p(16)\Big{]} ,
\label{Pu} \\ [5pt]
\mathcal{P}_p^{d \bar d} &=& \frac{1}{3} \Big{[}P_p(3)+P_p(5)+P_p(9)+P_p(12)+P_p(16)\Big{]}	\nonumber \\
	               &+& \Big{[}P_p(1)+P_p(8)+P_p(15)\Big{]} .
\label{Pd}
\end{eqnarray}
For the $s \bar {s}$ component, the probability is obtained by summing up linearly the relevant nonvanishing
contributions,
\begin{equation}
\mathcal{P}_p^{s \bar s} = \sum_{i=1}^{2}P_p(i) + P_p(4 ) + \sum_{i=6}^{8}P_p(i) + \sum_{i=10}^{11}P_p(i)+ \sum_{i=13}^{15}P_p(i) + P_p(17 ).
\label{Ps}
\end{equation}

The above probabilities can be related to the  ratios of interest in the present work, namely,
light quark--antiquark ratio ($r_\ell$), the strange sea suppression factor ($r_s$), and the strangeness
content of the proton ($\kappa_s$)
\begin{eqnarray}
r_\ell  &=&  \frac { P_{p}^{u\bar{u}} } { P_{p}^{d\bar{d}} },
\label{rl}   \\ [5pt]	
r_s  &=&  \frac { P_{p}^{s\bar{s}} } { P_{p}^{d\bar{d}} },
\label{rs}   \\ [5pt]
\kappa_s &=&  \frac { 2 P_{p}^{s \bar{s}} }  {P_{p}^{u\bar{u}} + P_{p}^{d \bar{d}} } .
\label{kappa}
\end{eqnarray}
%
%
\section{Results and Discussion}
\label{sec:Res}
In this section  we report our numerical results for the probabilities of five--quark states in the proton and the ratios,
Eqs.~(\ref{rl}) to (\ref{kappa}),
followed by comparisons with the CLAS~\cite{Park:2014zra} data and the outcomes of calculations performed by other
authors~\cite{Santopinto:2016fgs,Chang:2014lea,Ball:2012cx}.
\subsection{Results for quark--antiquark pair probabilities}
\label{sec:Rproba}

As described in~\cite{An:2012kj}, we only need to consider the five--quark configurations with $n_{r}=0$ and $l=1$.
Consequently, there are 17 different configurations which can be classified in four categories according to
the orbital and spin wave functions of the four--quark subsystem; the corresponding configurations are listed in Table~\ref{prob},
second column, using the shorthand notation for Young tableaux, where the subscripts $X$, $F$ and $S$ represent orbital, flavor and
spin, respectively.
%
%
\begin{table}[h!]
\caption{\footnotesize Predictions for probabilities of light and strange  five--quark configurations for the proton.
\label{prob}}
\scriptsize
\begin{tabular}{rrlccc}
\hline\hline
i && ~~~Category                  & $P_{p}^{u\bar{u}} $ & $ P_{p}^{d\bar{d}} $  & $P_{p}^{s\bar{s}} $   \\
 &&  configuration                  &  &   &    \\
\hline
     && ~~~~I) $[31]_{X}[22]_{S}$:                        &               &                    &                  \\
  1 &&  $[31]_{X}[4]_{FS}[22]_{F}[22]_{S}$          & 0            & 0.146(15)  & 0.010(1)    \\
  2 &&  $[31]_{X}[31]_{FS}[211]_{F}[22]_{S}$      & 0            & 0                & 0.004(1)    \\
  3 &&  $[31]_{X}[31]_{FS}[31]^1_{F}[22]_{S}$   & 0.011(1) & 0.005(1)  & 0    \\

  4 &&  $[31]_{X}[31]_{FS}[31]^2_{F}[22]_{S}$   & 0            & 0                &  0.003(1)    \\

   &&  ~~Total category I)                                        & 0.011(1) & 0.151(16) &  0.017(3)    \\
%
%
  && ~~~~II) $[31]_{X}[31]_{S}$: & && \\
  5 &&  $[31]_{X}[4]_{FS}[31]^1_{F}[31]_{S}$    & 0.048(5) & 0.024(3)  &   0  \\
  6 &&   $[31]_{X}[4]_{FS}[31]^2_{F}[31]_{S}$   & 0                & 0                 & 0.006(1)     \\
  7 &&  $[31]_{X}[31]_{FS}[211]_{F}[31]_{S}$    & 0                & 0                 & 0.003(1)    \\
  8 &&  $[31]_{X}[31]_{FS}[22]_{F}[31]_{S}$      & 0                & 0.006(1) & 0.002(0)    \\
  9 &&  $[31]_{X}[31]_{FS}[31]^1_{F}[31]_{S}$  & 0.003(0) & 0.002(1) &   0   \\
 10 &&  $[31]_{X}[31]_{FS}[31]^2_{F}[31]_{S}$ & 0                & 0                 &  0.001(0)   \\
   &&  ~~Total category II)                                     & 0.051(5) & 0.032(3)  &  0.012(1)    \\
%
%
  && ~~~~III) $[4]_{X}[22]_{S}$: & && \\
 11 &&  $[4]_{X}[31]_{FS}[211]_{F}[22]_{S}$     & 0                 & 0                &  0.009(0)    \\
 12 &&  $[4]_{X}[31]_{FS}[31]^1_{F}[22]_{S}$  & 0.028(2)  & 0.014(1) & 0   \\
 13 &&  $[4]_{X}[31]_{FS}[31]^2_{F}[22]_{S}$  & 0                  & 0                & 0.007(1)    \\
   &&  ~~Total category III)                                & 0.028(2)  & 0.014(1)  &  0.016(1)    \\
%
%
  && ~~~~IV) $[4]_{X}[31]_{S}$: & && \\
 14 &&  $[4]_{X}[31]_{FS}[211]_{F}[31]_{S}$    & 0                & 0                    & 0.008(1)    \\
 15 &&  $[4]_{X}[31]_{FS}[22]_{F}[31]_{S}$      & 0                & 0.015(2)    & 0.004(1)   \\
 16 &&  $[4]_{X}[31]_{FS}[31]^1_{F}[31]_{S}$  & 0.008(1) & 0.004(0)    & 0     \\
 17 &&  $[4]_{X}[31]_{FS}[31]^2_{F}[31]_{S}$  & 0                 & 0                   & 0.002(0)     \\
   &&  ~~Total category IV)                               & 0.008(1) & 0.019(2)  &  0.014(2)    \\
   && Total all configurations                             & 0.098(10) & 0.216(22) &  0.057(6)    \\
\hline
\hline
\end{tabular}
\end{table}

The probabilities for $u\bar{u} $, $ d\bar{d} $ and $s\bar{s}$
per configuration and intervening in Eqs.~(\ref{Pu}) to (\ref{Ps}) are given in Table~\ref{prob}, columns 3 to 5.
Those probabilities, as well as  the ones for ${c\bar{c}}$~\cite{Duan:2016rkr}, were also used to compute the
normalization factor in Eq.~(\ref{norm}).

Extensive comparisons with the outcomes of other approaches for $P_{p}^{Q\bar{Q}}$,
$Q\bar{Q} \equiv u\bar{u}, d\bar{d}, s\bar{s}, c\bar{c}$ were reported in~\cite{An:2012kj,Duan:2016rkr}
and  led in general to compatibility of our results with those achieved by other authors.
As documented in~\cite{An:2012kj} the free parameters of our model are taken from the literature, except one of them.
This latter, a common factor of the matrix elements of the transitions between three- and five--quark components,
was found~\cite{Duan:2016rkr} to be $V$ = 572 $\pm$ 47 MeV,  by successfully fitting the experimental
data~\cite{Towell:2001nh} for the proton flavor asymmetry
$\mathcal{A}_p=\bar{d} - \bar{u} \equiv {\mathcal{P}_p^{d \bar d}} - {\mathcal{P}_p^{u \bar u}}=0.118\pm0.012$.
The only source of uncertainty in the probabilities (Table~\ref{prob}),  comes from that  factor.
It is worthy to note that for the ratios in Eqs.~(\ref{rl}) to (\ref{kappa}) the common factor $V$ divides out.
Accordingly, no parameters were adjusted in the frame of the present work.


\subsection{Results for ratios and comparisons with data and other approaches}
\label{sec:Rratio}
Using a 5.5 GeV electron beam at Jefferson Laboratory (JLab), the CLAS Collaboration measured~\cite{Park:2014zra} the ratios
of pseudoscalar mesons electroproduction ($e p \to e^{\prime} K^+  \Lambda ,~ e^{\prime} \pi^ \circ  p,~e^{\prime} \pi^+ n$)
exclusive reaction cross sections in the phase space kinematics  covering $W$ = 1.65 -- 2.55 GeV and  $Q^2$ = 1.6 -- 4.6 GeV$^2$.
Ratios of the measured final state meson-nucleon cross sections were then related to the ratios of quark--antiquark pairs,
Eqs.~(\ref{rl}) and (\ref{rs}),  {\it via}  a simple model of pair creation on one of the quarks of the proton target, supposed to be
exclusively a three-quark state.
The extracted ratios~\cite{Park:2014zra} are
\begin{eqnarray}
r_\ell  &=& \frac {u\bar{u}} {d\bar{d}} \approx 2\Big (\frac {\langle  \pi^ \circ  p \rangle} {\langle  \pi^ +  n \rangle}
             - \frac {1}  {16}   \Big ),
\label{rlexp}	\\ [5pt]
r_s  &=&  \frac {s\bar{s}} {d\bar{d}}   \approx \frac {\langle  K^+  \Lambda \rangle} {\langle  \pi^ +  n \rangle} .
\label{rsexp}
\end{eqnarray}
Note that the strangeness content of the proton can be expressed in terms of $r_\ell $ and $r_s$
\begin{equation}
\kappa_s =   \frac { 2 s\bar{s} } {  u\bar{u}+ d\bar{d} }= \frac { 2r_s }  {r_\ell + 1 } .
\label{kappaexp}
\end{equation}

The CLAS data for $r_\ell $ and $r_s$~\cite{Park:2014zra}, as well as the extracted value  for
$\kappa_s$~\cite{Santopinto:2016fgs} are given in Table~\ref{comp} (last row) and compared with the predictions of
our approach and the outcomes from other investigations~\cite{Santopinto:2016fgs,Chang:2014lea,Ball:2012cx}.
To our knowledge, this set of data constitutes the first experimental results on {\it both}  light and strange  quark--antiquark
ratios, albeit with model dependent extraction and rather large uncertainties
($\delta r_\ell$ = 24\%, $\delta r_s$ =  32\%, $\delta \kappa_s$ = 36\%), dominated by systematic errors
($\delta_{sys.}/\delta_{stat.} \approx$ 7).
Note that $\delta_{sys.}$ comes from the experimental uncertainties and do not include the ones
due to the simple semiclassical model used in extracting the quark--antiquark ratios from the measurement.
%
%
\begin{table}[t]
\caption{\footnotesize The probabilities of quark--antiquark pairs and their ratios; see Eqs. (9) to (16).
\label{comp}}
%
%
\begin{tabular}{lccccccccc}
\hline\hline
%
 Reference & Approach && $P_{p}^{u\bar{u}} $&$P_{p}^{d\bar{d}} $&$P_{p}^{s\bar{s}}$&  & $r_\ell$  & $r_s$ &$\kappa_s$  \\
\hline
{Present work}                                                  & E$\chi$CQM  && 0.098(10) & 0.216(22)&  0.057(6) & & 0.45 & 0.26 &  0.36  \\
Santopinto {\it et al.} \cite{Santopinto:2016fgs}  &UQM              &&           &            &                                                && 0.57 & 0.26 &  0.34  \\
%
%
Chang-Peng  \cite{Chang:2014lea}
%
                                                                              &  BHPS (S1)&& 0.194 & 0.312  & 0.111& & 0.62 & 0.36  &  0.44  \\
%
%
                                                                            &BHPS (S3) && 0.213 & 0.331 & 0.039 & & 0.64 & 0.12    &  0.14  \\
 %
 %
  Ball  {\it et al.} \cite{Ball:2012cx}                     & NNPDF2.3 noLHC  &&  &   &  &                                  & & 0.39(10) &  0.30(9)  \\
                                                                         & NNPDF2.3 LHC  &&  &   &  &                                      & & 0.43(11) &  0.35(9)  \\
Mestayer {\it et al.} \cite{Park:2014zra}               & CLAS Data   && &&& &0.74(18) & 0.22(7)& 0.25(9)   \\
 \hline
\hline
\end{tabular}
\end{table}
%
%

The predictions of our model (Table~\ref{comp}, row 2) embodying only the nonperturbative mechanism due to the intrinsic
quark--antiquark pairs account for roughly
60\% of $r_\ell$, underestimating the measured central value by 1.6$\sigma$.
However, our model reproduces  $r_s$ and $\kappa_s$ within $\approx$1$\sigma$.
A plausible explanation would be that at the CLAS kinematics, the probabilities of perturbative  production of light quark--antiquark pairs
are larger than that for the $s\bar{s}$ ones; still dominated by the nonperturbative mechanisms.
Note that our results come from probabilities including all 17 configurations  (last row in Table~\ref{prob}).
We checked ratios per category (Table~\ref{prob}, rows 8, 16, 21 and 27), but none of them improved the predictions for ratios,
endorsing that any configuration--truncated set leads to unrealistic results~\cite{An:2012kj,An:2013daa,Duan:2016rkr,An:2014aea}.

In the following we proceed to comparisons among various phenomenological results (Table~\ref{comp}, rows 3 to 7)  and data
(last row).

Interpreting the CLAS data, Santopinto~{\it et al.} \cite{Santopinto:2016fgs} performed a calculation within the Unquenched Quark Model
(UQM), based on a quark model with continuum components, to which quark--antiquark pairs  are added perturbatively employing a
$^3P_0$ model.
Their results referring to the $Q\bar{Q}$ production ratios with pseudoscalar mesons in combination with octet and decuplet baryons
are given in Table~\ref{comp}, row 3.
The experimental  value for $r_\ell$ is reproduced within 1$\sigma$, while for  $r_s$ and $\kappa_s$ their results are comparable
with ours.

Chang and Peng~\cite{Chang:2011a,Chang:2011b,Chang:2014lea} investigated the  intrinsic ${Q\bar{Q}} $ states in the proton by
generalizing the BHPS model, as mentioned in the Introduction.
 In their most recent work~\cite{Chang:2014lea}, the authors perform a comprehensive study of the  latest results from the HERMES
 Collaborations~\cite{Airapetian:2008qf,Airapetian:2012ki,Airapetian:2013zaw}.
 The most recent experimental data~\cite{Airapetian:2013zaw} are then classified~\cite{Chang:2014lea} in three sets, for which
$P_{p}^{u\bar{u}} $,  $P_{p}^{d\bar{d}} $ and $P_{p}^{s\bar{s}}$ are extracted  by evolving the light-cone five--quark  BHPS model to
$Q^2$ = 2.5 GeV$^2$  for the initial scale  values $\mu$ = 0.3 and 0.50 GeV.
 In Table~\ref{comp} (rows 4 and 5), their results for  two of the sets  ($S_1$ and $S_3$)
with $P_{p}^{s\bar{s}} \neq 0$ are reported, for $\mu$ = 0.3 GeV; where $r_\ell$,  $r_s$, and $\kappa_s$ were computed
following Eqs.~(\ref{rl}) to (\ref{kappa}).
First, we focus on the light quarks' results, for which the probabilities determined within the BHPS model, turn out to be larger than
our predictions and their $r_\ell$ value approaches the experimental data within better than 1$\sigma$.  For the strangeness sector the situation is more contrasted, with $P_{p}^{s\bar{s}}$, and $r_s$ and
 $\kappa_s$ varying by roughly a factor of 3 between $S1$ and $S3$.
Both sets show larger deviation from the data than our predictions
for $r_s$ and $\kappa_s$. In the BHPS based approach, these latters are overestimated by roughly 2$\sigma$ in $S1$ and
underestimated by about 1$\sigma$ in $S3$.
The BHPS results  for the initial scale value $\mu$ = 0.50 GeV show comparable trends, although the five--quark probabilities
turn out to be $\approx$50\% smaller than those for $\mu$ = 0.30 GeV.

 An extensive study to determine $r_s$ and  $\kappa_s$ was performed by the NNPDF Collaboration~\cite{Ball:2012cx}.
The main idea of this approach~\cite{DelDebbio:2004xtd} is to train a set of neural networks on a set of Monte Carlo replicas of the experimental
data reproducing their probability distribution.
Accordingly, Ball  {\it et al.} \cite{Ball:2012cx}  proceeded through global fits to extended sets of data obtained from electro- and hadro-production
processes; in particular, deep inelastic scattering, Drell-Yan, gauge boson  and jet production (see Table 7 in \cite{Ball:2012cx} for
 relevant references to some forty data sets).
Concerning the quantities of interest in the present work, the NNPDF Collaboration extracted the strangeness  and strangeness momentum
 fractions  via the following expressions:
\begin{eqnarray}
r_s (Q^2) &=& \frac {\int_{0}^{1}  x[s(x, Q^2)  +  \bar{s} (x, Q^2)  ] dx }
{ 2 \int_{0}^{1}  x \bar{d} (x, Q^2)  dx }  ,
\label{lams}	\\ [8pt]
\kappa_s (Q^2)&=&  \frac {\int_{0}^{1}  x[s(x, Q^2)  +  \bar{s} (x, Q^2)  ] dx }
{  \int_{0}^{1}  x [ \bar{u} (x, Q^2)  + \bar{d} (x, Q^2)  ] dx }.
\label{kaps}
\end{eqnarray}

\vspace{0.2cm}

The fitted data span a large domain in the Bjorken scaling variable $x$.
For small values of $x$ the quark--antiquark production process  is due to perturbative phenomena arising from extrinsic (e.g. $g \to q \bar{q}$)
components, but in the range of $ 0.2 \lesssim x \lesssim 0.8$ contributions from the intrinsic quark--antiquark pairs become the dominant
mechanism.
The outcomes of that work, at $Q^2$=2 GeV$^2$, without and with the LHC data,  are given in Table~\ref{comp} (rows 6 and 7) and do
not produce drastic changes arising from the LHC data.
Comparing results from the NNPDF Collaboration with ours shows that the agreement between the two approaches is within less than
1.5$\sigma$ for $r_s$ and better than 1$\sigma$ for $\kappa_s$.
Compilation of the extracted values for $\kappa_s$ from experimental data on neutrino induced opposite-sign dimuon events (Table 4
 in~\cite{Chang:2014jba}) leads to the range $\kappa_s$ = 0.33 -- 0.59.
Interestingly, the two extreme values result from the latest measurements:  0.33~$\pm$~0.07 from CHORUS~\cite{KayisTopaksu:2008aa}
and 0.59~$\pm$~0.02 from NOMAD~\cite{Samoylov:2013xoa} Collaborations.
While the former one is in the range of the values from the phenomenological approaches (Table~\ref{comp}),  the latter one turns out to
be significantly larger than those findings.
As emphasized by Chang and Peng~\cite{Chang:2014jba}, the extracted values from experiments depend on the order of perturbative QCD
corrections employed.
Actually, such trends are extensively illustrated based on recent developments in the determination of PDFs in global QCD analyses
~\cite{McNulty:2016xtv}, results from various approaches ~\cite{Accardi:2016ndt} and the impact of different
 data sets on the extracted PDFs~\cite{Alekhin:2014sya,Accardi:2016qay}.


\section {Summary and conclusions}
\label{sec:conclu}
In the present work, we investigated the recently measured~\cite{Park:2014zra} quark--antiquark ratios
$r_\ell={u\bar{u}} / {d\bar{d}}$, $r_s={s\bar{s}} / {d\bar{d}}$, and
$\kappa_s=2{s \bar{s}} / ({u\bar{u}} + {d \bar{d}})$,
attempting to single out the role of the intrinsic ${Q\bar{Q}}$ components in the proton's wave function, with
${Q\bar{Q}} \equiv {u\bar{u}}, {d\bar{d}}, {s\bar{s}}, {c\bar{c}}$.
For that purpose, we employed the recently developed extended chiral constituent quark
model~\cite{An:2012kj,An:2013daa,An:2014aea,Duan:2016rkr}; within which the baryons are considered as admixtures of
three- and five--quark states.
Probabilities of the five--quark components were calculated using the $^{3}P_{0}$ transition
operator~\cite{Le Yaouanc:1972ae}.
The quark--antiquark pair probabilities were determined by fixing a common factor of the matrix elements of the transitions between
three-- and five--quark components~\cite{Duan:2016rkr} by fitting the experimental data for the proton flavor asymmetry
$\mathcal{A}_p=\bar{d} - \bar{u} \equiv {\mathcal{P}_p^{d \bar d}} - {\mathcal{P}_p^{u \bar u}} = 0.118\pm0.012$
~\cite{Towell:2001nh}.
However, that factor divides out in the studied ratios [Eqs.~(\ref{rl}) to (\ref{kappa}].
Accordingly, our predictions for the ratios were obtained without any adjusted parameters on the CLAS data~\cite{Park:2014zra}.
Moreover, the set of parameters taken from the literature~\cite{An:2012kj}, and utilized in the present work,
allowed us predicting successfully the strangeness magnetic form factor of the proton~\cite{An:2013daa} and producing results compatible with
findings within other formalisms for the sigma terms:
$\sigma_{\pi N}$, $\sigma_{s N}$~\cite{An:2014aea}, and $\sigma_{c N}$~\cite{Duan:2016rkr}.

The same flavor asymmetry $\mathcal{A}_p$ data was fitted also by Chang and Peng~\cite{Chang:2011a,Chang:2011b,Chang:2014lea}
 within a generalized BHPS model.
However, their extracted probabilities for $\mathcal{P}_p^{u \bar {u}}$ and $\mathcal{P}_p^{d \bar {d}}$ differ significantly from
ours (Table~\ref{comp}) by a factor of 2 and 50\%, respectively.
Accordingly, $r_\ell$ turns out to be 50\% higher in the BHPS model~\cite{Chang:2014lea} than in ours.
So, the $\mathcal{A}_p$ data does not put strong enough constraints on the models.
The two values for ${\mathcal{P}_p^{s \bar s}}$ in the BHPS approach~\cite{Chang:2014lea} show variation by a factor of 3
and our prediction falls in between; that is also the case comparing the two sets' predictions ($S1$ and $S3$) with ours for
 $r_s$ and $\kappa_s$.

The two other phenomenological works~\cite{Santopinto:2016fgs,Ball:2012cx} discussed in this paper, embody contributions from
both intrinsic and extrinsic higher Fock states, especially in the case of the NNPDF approach~\cite{Ball:2012cx}.
The UQM model's values~\cite{Santopinto:2016fgs} compared with ours suggest that the intrinsic component accounts for
roughly 80\% in $r_\ell$ and almost 100\% in $r_s$ and $\kappa_s$.
In other words, the CLAS data~\cite{Park:2014zra} for strangeness are dominated by the intrinsic five--quark states.
The situation is different with respect to the NNPDF Collaboration findings~\cite{Ball:2012cx} due to the fact
that their fitting runs over a large range in Bjorken-$x$, including (very) low--$x$ region, dominated by perturbative mechanisms.
Then,  $r_s$ turns out to be more sensitive than $\kappa_s$ to that latter effect.
However, both $r_s$ and $\kappa_s$ are compatible with the CLAS data and our predictions, within the reported uncertainties.

In summary,
i) from theory--experiment comparisons performed within the present work we infer that the CLAS data could be interpreted as receiving
contributions from both intrinsic and, to a lesser extent, from extrinsic ${Q\bar{Q}}$ components, while the ${s \bar{s}}$ pairs are
mainly from nonperturbative origin;
ii) the present status of a rather large number of data sets does not allow sharp comparisons with various
phenomenological approaches, showing the need for more accurate measurements and their extension to medium and high Bjorken-$x$ regions
($x \gtrsim$  0.1).

Actually, the ongoing SeaQuest experiment~\cite{Reimer:2016dcd}, measuring Drell-Yan scattering in Fermilab, aims at providing more
precise data on light quark--antiquark components, extending the Bjorken-$x$ domain to $x \approx$ 0.45, where the sea quark distributions
are dominated by nonperturbative regime.
Moreover, determination of the PDFs will benefit from the upcoming data from facilities such
as the LHC~\cite{Ball:2014uwa,Alekhin:2017kpj}, JLab~\cite{Montgomery:2017qrz}, J-PARC~\cite{Kumano:2015gna} and
NICA~\cite{Musulmanbekov:2011zz,Brodsky:2016tew}.
Finally, progress in the lattice QCD calculations~\cite{Liu:2016djw,Alexandrou:2016bud} appears very promising in pinning down
 the genuine quark--antiquark pairs quest in the proton.
We might then expect achieving in the near future a comprehensive understanding of the role and importance of the intrinsic five--quark
components in baryons.

%
\begin{acknowledgments}
We are grateful to Mac Mestayer for valuable clarifications on the CLAS data.
This work is partly supported by the National Natural Science Foundation of China under Grant
No. 11675131, Chongqing Natural Science Foundation under Grant
No. cstc2015jcyjA00032, and Fundamental Research Funds for the Central Universities under Grant No. SWU115020.
\end{acknowledgments}
%
%
%
%

%
\end{document}